\documentclass{moriond}

\usepackage{amsfonts}
\usepackage{dcolumn}
\usepackage{bm}
\usepackage{amsmath}
\usepackage{nicefrac}
\usepackage{amsthm}
\usepackage{amssymb}
\usepackage{graphicx}
\usepackage{color}
\usepackage{url}
\usepackage{cancel}
\usepackage{units}
\usepackage{upgreek}
\usepackage[shortcuts]{extdash}
\usepackage[normalem]{ulem}
\usepackage[export]{adjustbox}
\usepackage{floatflt}
\usepackage{placeins}
\usepackage{wrapfig}
\usepackage{hyperref} % Required for adding links	and customizing them
\definecolor{linkcolour}{rgb}{0.,0.,0.8} % Link color
\hypersetup{colorlinks=true,urlcolor=linkcolour,linkcolor=linkcolour,citecolor=linkcolour} % Set link colors throughout the document

\renewcommand{\v}[1]{\boldsymbol{#1}}		%bold-math for vectors

\newcommand{\Photo}{}

\begin{document}
\vspace*{4cm}
\title{Hunting for Axionlike Dark Matter by Searching for an Oscillating Neutron Electric Dipole Moment}

\author{ Nicholas J. Ayres for the nEDM experiment at PSI }
\address{Department of Physics and Astronomy, University of Sussex, Falmer,\\ Brighton BN1 9QH, United Kingdom}

\maketitle \abstracts {We report on a search for ultra-low-mass axion-like dark matter by analysing the ratio of the spin-precession frequencies of stored ultracold neutrons and $^{199}$Hg atoms for an axion-induced oscillating electric dipole moment of the neutron and an axion-wind spin-precession effect. No signal consistent with dark matter is observed for the axion mass range $10^{-24}~\textrm{eV} \le m_a \le 10^{-17}~\textrm{eV}$. Our null result sets the first laboratory constraints on the coupling of axion dark matter to gluons, which improve on astrophysical limits by up to 3 orders of magnitude,  and also improves on previous laboratory constraints on the axion coupling to nucleons by up to a factor of 40. The results were initially presented in Phys. Rev. X \textbf{7}, 041034\cite{Abel2017}, of which this proceeding is largely a summary.}

%%%%%%%%%%
\section{Introduction}
\label{Sec:Introduction}

Astrophysical and cosmological observations indicate that $26 \%$ of the total energy density and $84 \%$ of the total matter content of the Universe is dark matter (DM) \cite{Planck2015}, the identity and properties of which still remain a mystery. 
One of the leading candidates for cold DM is the axion.

It is reasonable to expect that axions interact non-gravitationally with standard-model particles. 
Direct searches for axions have thus far focused mainly on their coupling to the photon (see the review \cite{Axion-Review2015} and references therein). 
Recently, however, it has been proposed to search for the interactions of the coherently oscillating axion DM field with gluons and fermions, which can induce oscillating electric dipole moments (EDMs) of nucleons \cite{Graham2011} and atoms \cite{Stadnik2014A,Roberts2014A,Roberts2014B}, and anomalous spin-precession effects \cite{Stadnik2014A,Graham2013}. 
The frequency of these oscillating effects is dictated by the axion mass, and more importantly, these effects scale linearly in a small interaction constant \cite{Graham2011,Stadnik2014A,Roberts2014A,Roberts2014B,Graham2013}, whereas in previous axion searches, the sought effects scaled quadratically or quartically in the interaction constant \cite{Axion-Review2015}. 

In the present work, we focus on the axion-gluon and axion-nucleon couplings
\begin{align}
\label{Axion_couplings}
\mathcal{L}_{\textrm{int}} = \frac{C_G}{f_a} \frac{g^2}{32\pi^2} a G^{b}_{\mu \nu} \tilde{G}^{b \mu \nu}  - \frac{C_N}{2f_a} \partial_\mu a ~ \bar{N} \gamma^\mu \gamma^5 N \, ,
\end{align}
where $G$ and $\tilde{G}$ are the gluonic field tensor and its dual, $b=1,2,...,8$ is the  color index, $g^2 / 4 \pi$ is the color coupling constant, {\color{black}$N$ and $\bar{N} = N^\dagger \gamma^0$ are the nucleon field and its Dirac adjoint,} $f_a$ is the axion decay constant, and $C_G$ and {\color{black}$C_N$} are model-dependent dimensionless parameters. 

The axion-gluon coupling in (\ref{Axion_couplings}) induces the following oscillating EDM of the neutron via a chirally-enhanced 1-loop process~\cite{Witten1979,Pospelov1999}
\begin{equation}
\label{eq:nEDM_axion}
d_\mathrm{n}(t) \approx +2.4 \times 10^{-16} ~ \frac{C_G a_0}{f_a} \cos(m_a t) ~ e \cdot \textrm{cm} \, .
\end{equation}
The amplitude of the axion DM field, $a_0$, is fixed by the relation $\rho_a \approx m_a^2 a_0^2 /2$. 
In the present work, we assume that axions saturate the local cold DM energy density $\rho_{\rm DM}^{\rm local} \approx 0.4~\textrm{GeV/cm}^3$ \cite{Catena2010}. 
The derivative coupling of an oscillating galactic axion DM field, $a = a_0 \cos(m_a t - \v{p}_a \cdot \v{r})$, with spin-polarized nucleons in (\ref{Axion_couplings}) induces time-dependent energy shifts according to
\begin{equation}
\label{potential_axion-wind}
H_{\textrm{int}} (t) = \frac{C_N a_0}{2 f_a} \sin(m_a t) ~ \v{\sigma}_N \cdot \v{p}_a \, .
\end{equation}
The term $\v{\sigma}_N \cdot \v{p}_a$ adds a dependence on the relative direction of the motion of the axion field and the spin quantization axis, causing additionally a daily modulation in precession frequency as the earth rotates, which we do not consider in this analysis.

Here, we report on a search for an axion-induced oscillating EDM of the neutron (nEDM) based on an analysis of the ratio of the spin-precession frequencies of stored ultracold neutrons and $^{199}$Hg atoms, which is a system that had previously also been used as a sensitive probe of new non-EDM physics \cite{Altarev2009,Altarev2010,Afach2015_NF}. 
We divided our analysis into two parts. 
We first analyzed the Sussex--RAL--ILL nEDM experiment data \cite{Baker2014}, covering oscillation periods longer than days (\emph{long time--base}). 
Then we extended the analysis to the data of the PSI nEDM experiment \cite{Baker2011}, which allowed us to probe oscillation periods down to minutes (\emph{short time--base}). 
Our analysis places the first laboratory constraints on the axion-gluon coupling.
We also report on a search for an axion-wind spin-precession effect, using the data of the PSI nEDM experiment. Our analysis places the first laboratory constraints on the axion-nucleon coupling from the consideration of an effect that is linear in the interaction constant.

\section{Long time-base analysis}
The Sussex--RAL--ILL room temperature nEDM experiment ran from 1998 to 2002 at the PF2 beamline at the Institut Laue-Langevin (ILL) in Grenoble, France. 
This experiment set the current world-best limit on the permanent time-independent neutron EDM, published in 2006 \cite{Baker2006}. The data were subsequently reanalyzed to give a revised limit in 2015 \cite{Pendlebury2015}. The technical details of the apparatus are described in full in \cite{Baker2014}, but we summarize the main experimental details here for the reader.

The experiment was based on Ramsey interferometry \cite{Ramsey1950} of ultracold neutrons. The neutrons were stored in parallel or antiparallel electric and magnetic fields, where their Larmor precession frequency is given by
\begin{equation}
\label{eq:Larmor}
h\nu_\mathrm{n} = 2 \left| \mu_\mathrm{n} B \pm d_\mathrm{n} E \right| \, ,
\end{equation}
with the sign depending on the field configuration. 
$E$ and $B$ are the magnitudes of the electric and magnetic fields, respectively. 
By measuring the frequency difference between the two field configurations, a value for the neutron EDM, $d_\mathrm{n}$, was inferred. The measurement was conducted in a series of \emph{cycles}, each approximately 5 minutes long. A cycle began with a filling of neutrons polarized along the fields into the precession chamber from the ultracold neutron source~\cite{Steyerl1986}, and yields a single estimate of the neutron precession frequency. We name a series of consecutive cycles taken over typically 1-2 days in an identical magnetic field configuration, but the direction of the electric field periodically reversed, a \emph{run}.

In order to suppress cycle--to--cycle changes in the magnetic field, the analysis was performed on the ratio of the neutron and mercury precession frequencies $R$, which, using (\ref{eq:Larmor}), is \cite{Baker2014}
\begin{align}
\label{eq:R}
R &\equiv \frac{\nu_\mathrm{n}}{\nu_\textrm{Hg}} = \frac{\mu_\mathrm{n}}{\mu_\textrm{Hg}} \pm \left( d_\mathrm{n} - \frac{\mu_\mathrm{n}}{\mu_\textrm{Hg}} \, d_\textrm{Hg} \right) \frac{2 E}{ h  \nu_\textrm{Hg}} + \Delta \, ,
\end{align}
where the signs correspond to parallel and antiparallel field configurations. $\Delta$ encapsulates all higher-order terms and systematic effects, which are corrected for when a run is analyzed~\cite{Pendlebury2015}.

In the case of the long time--base analysis, we considered the time series of $d_\mathrm{n}$ measurements from individual runs.

As an estimator of the power spectrum of the data, we used the Least Squares Spectral Analysis (LSSA) periodogram \cite{Scargle1982,Cumming2004}, where the amplitude at frequency $f$ was estimated by the amplitude of the best fit oscillation of that frequency. We evaluated the periodogram at a set of 1334 trial frequencies, evenly spaced between \unit[100]{pHz} (arbitrarily chosen, a period of about 300 years, much longer than the four--year span of the data set) and $\unit[10]{\upmu Hz}$ (a period of about a day, the time it typically took to get one $d_\mathrm{n}$ estimate).

\begin{figure}
	\centering
	%\begin{subfigure}
	\begin{minipage}[t]{0.45\textwidth}
	\includegraphics[width=\textwidth]{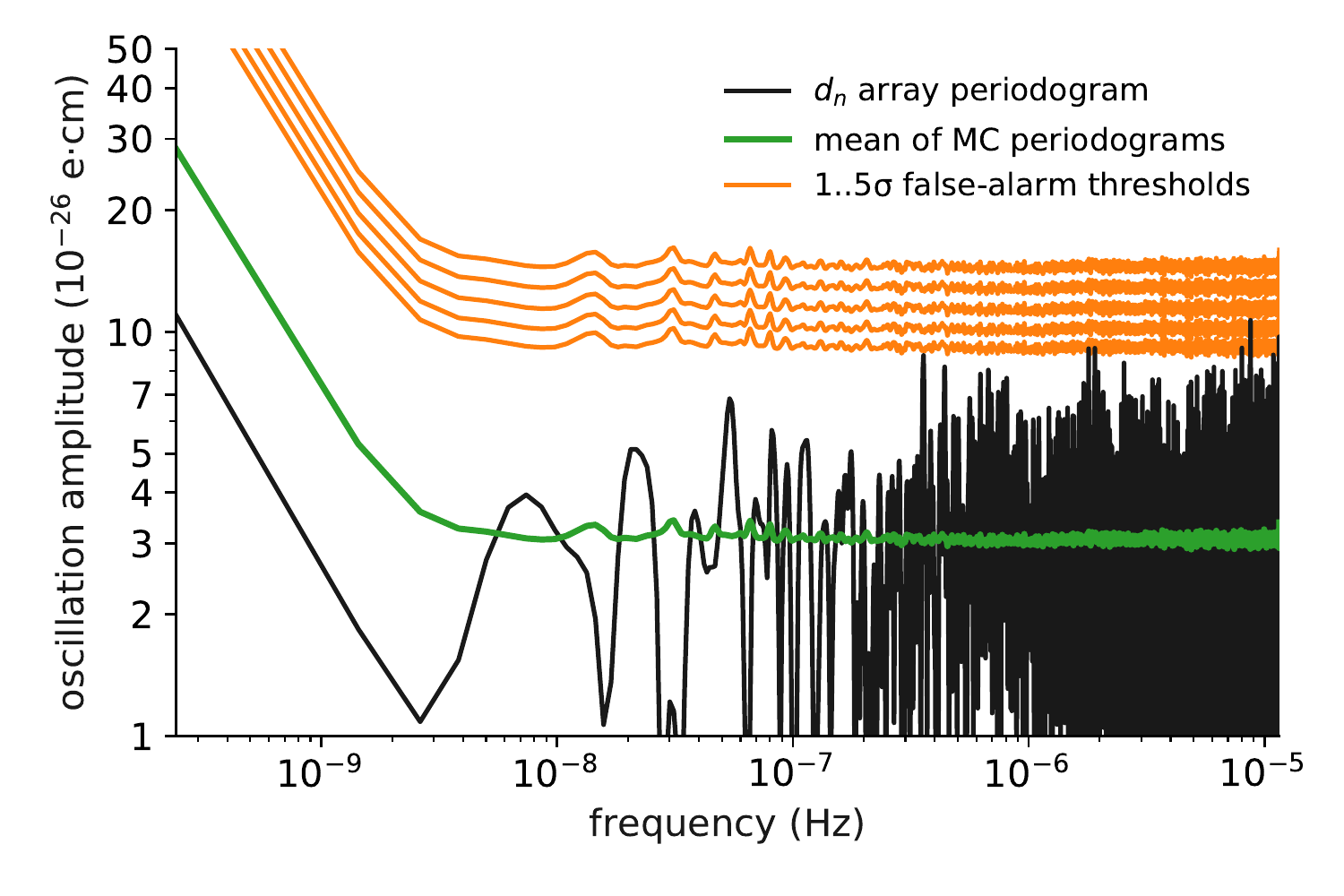}
	\caption{ \footnotesize{ The periodogram of the array of neutron EDM ($d_\mathrm{n}$) estimates from the ILL measurement (black line). 	The mean of Monte Carlo (MC)-generated periodograms, assuming no signal is present, is depicted in green. MC is used to deliver false--alarm thresholds (global $p$-values), marked in orange for $1,2,…,5\,\sigma$ levels (from bottom to top). The highest peak has the global $p$\-/value 0.53, consistent with a non-detection. Reprinted figure with permission from \protect\cite{Abel2017}. Copyright 2017 by the American Physical Society.}}
	\label{fig:ILL_detection}
	\end{minipage}
	%\end{subfigure}
	\qquad
	%\begin{subfigure}
	\begin{minipage}[t]{0.45\textwidth}
	\includegraphics[width=\textwidth]{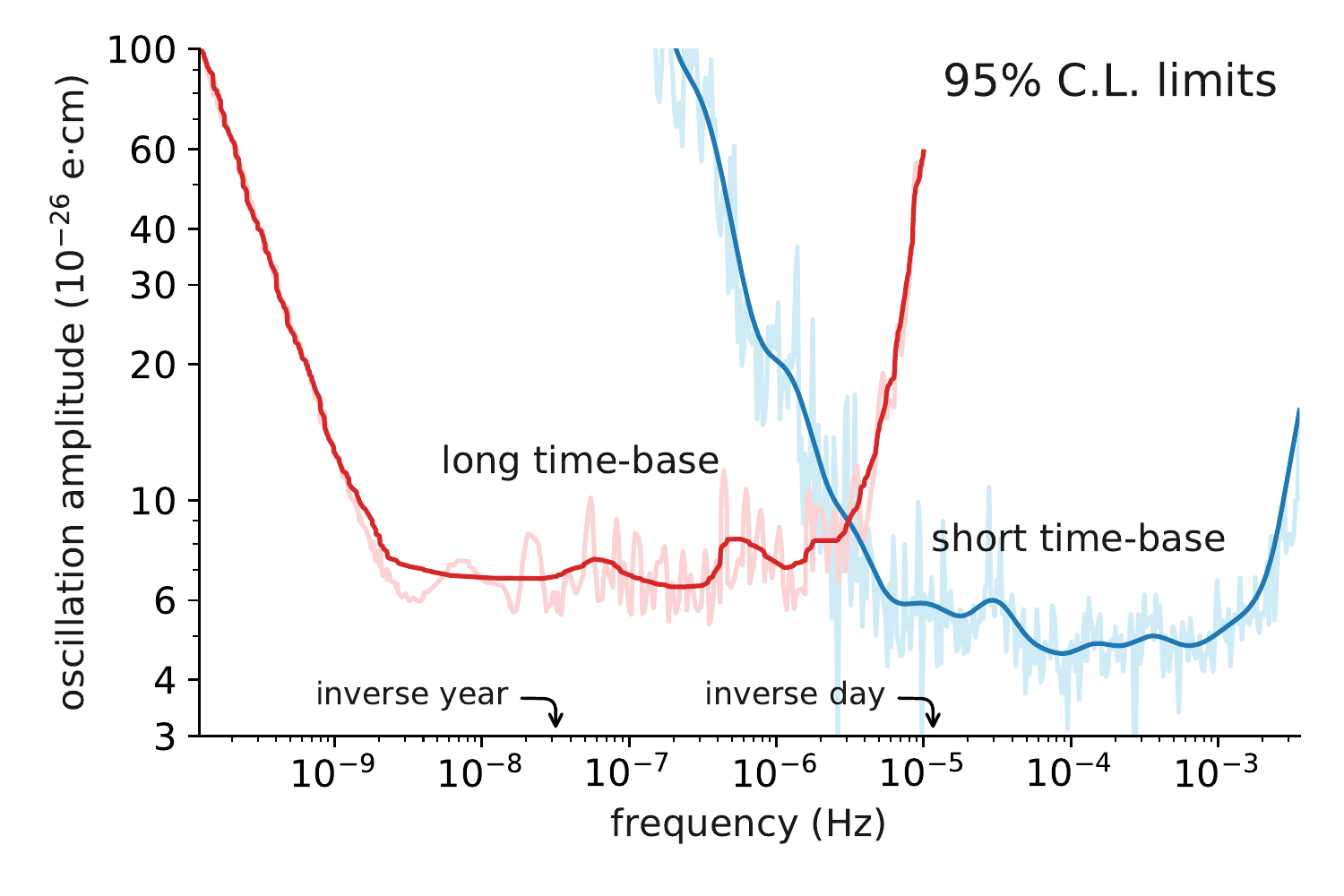}
	%\end{subfigure}
	\caption{ \footnotesize{The 95\% C.L. limits on the amplitude of oscillation in the quantity $d_\mathrm{n} - \left( \mu_\mathrm{n} / \mu_\textrm{Hg} \right) \, d_\textrm{Hg}$, as a function of frequency thereof. The limits from the long (ILL data) and short (PSI data) time--base analyses are depicted by the red and blue curves, respectively, with the area above these curves being excluded. The raw limits delivered by the analysis, with substantial noise, are depicted by the light lines, while the smoothed versions are given in bold. From \protect\cite{Abel2017}.}}
	\label{fig:ecm_limits}
\end{minipage}
\end{figure}

To obtain the expected distribution of the periodogram, we performed Monte Carlo (MC) simulations. At each frequency, we estimated the cumulative distribution function (CDF) of the LSSA power. Extreme events in the tails of the distribution are expensive to access directly with MC. For this reason, to the discrete CDF estimates we fitted, at each $i^\textrm{th}$ frequency, the functional form of the LSSA-power CDF~\cite{Scargle1982}
\begin{equation}
\label{eq:powerdistribution}
F_i(\mathcal{P}) = 1 - A_i\,\exp(-B_i\, \mathcal{P}) \, ,
\end{equation}
where $\mathcal{P}$ is the power, while $A_i$ and $B_i$ are fit parameters. The local $p$-values are given by
\begin{equation}
\label{eq:localpvalue}
p_{\mathrm{local}, i} = 1 - F_i(\mathcal{P}_i) \, ,
\end{equation}
where $\mathcal{P}_i$ is the LSSA power of the measured $d_\mathrm{n}$ time series at the $i^\textrm{th}$ frequency.
If the local $p$\-/values at different trial frequencies were uncorrelated, the global $p$\-/value would be given by \cite{Algeri2016}
\begin{equation}
\label{eq:pvalues}
p_\mathrm{global} = 1 - (1 - p_\mathrm{local})^N \, ,
\end{equation}
where $N$ is the number of trial frequencies. However, we did not need to make this assumption. Instead, we made use of the set of MC datasets.
In each, we found the minimal local $p$\-/value and estimated its CDF, assuming it has the form (\ref{eq:pvalues}), but left $N$ as a free parameter. We found the best fit value $N_\mathrm{effective} = 1026$. For each frequency, we marked the power necessary to reach the global $p$\-/values corresponding to $1,2,…,5\,\sigma$ levels as orange lines in Fig.\,\ref{fig:ILL_detection}. The minimal local $p$\-/value of the dataset translates to the global $p$\-/value of 0.53, consistent with a non-detection.

In order to obtain limits on the oscillation amplitude parameter, we again used MC simulations. We discretized the space of possible signals, spanned by their frequency and amplitude. We chose a sparser set of 200 frequencies, as we did not expect highly coherent effects in the sensitivity of detection. For each discrete point, we generated a set of 200 MC datasets containing the respective, perfectly coherent signal and assumed that the oscillation is averaged over the duration of the run.

For each fake dataset, we evaluated the LSSA amplitude only at the frequency of the signal and compared its distribution (extrapolating with the functional form of Eq.\,(\ref{eq:powerdistribution})) with the best-fit amplitude in the data and defined the $p$\-/value to be left--sided. We found the 95\% confidence-level exclusion limit as the 0.05 isocontour of the $\mathrm{CL}_s$ statistic \cite{PDG2016}. 
The limit is shown as the red curve in Fig.\,\ref{fig:ecm_limits}. 
We are most sensitive to periods shorter than the timespan of the dataset ($\sim 4$ years), but rapidly lose sensitivity for periods shorter than the temporal spacing between data points ($\sim 2$ days), since the expected signal would essentially average to zero over these short time scales. 

\section{Short time-base analysis}
In 2009, the Sussex--RAL--ILL apparatus was moved to the new ultracold neutron source at the Paul Scherrer Institute (PSI), Villigen, Switzerland~\cite{Anghel2009}, where a number of improvements were made~\cite{Baker2011,Afach2015USSA,Ban2016NANOSC}. In 2015, the apparatus was fully commissioned and began to take high-sensitivity EDM data. The whole data set, taken from August 2015 until the end of 2016, with a higher accumulated sensitivity than the ILL one, was considered in this analysis. For the PSI experiment's data, we performed a lower--level oscillation search on the array of $R$ measurements. Since an $R$ estimate was obtained every cycle ($\unit[\approx 300]{s}$), rather than every 1--2 days as for a $d_\mathrm{n}$ estimate, it has an increased sensitivity to higher frequencies. Additionally, the analysis could benefit from the addition of 16 atomic cesium vapor magnetometers \cite{Knowles2009}, located directly above and below the precession chamber (inside the electrodes).

The dominant time-dependent systematic effect, encapsulated in $\Delta$ of Eq.\,\eqref{eq:R}, would have given rise to non-statistical temporal fluctuations if not accounted for. Namely, $R$ is sensitive to drifts in the vertical gradients of the magnetic field. While the thermal mercury atoms filled the chamber homogeneously, the center of mass of the ultracold neutron population was lower by several millimeters~\cite{Afach2014magmoment,Afach2015,Pendlebury2015}. To compensate gradient drifts on a cycle-to-cycle basis, the vertical gradient was measured by fitting a second order parametrization of the magnetic field to the measurements of the cesium magnetometers. However, the absolute calibration of these sensors was insufficient to allow corrections of the deliberate, large changes between runs.

To account for this, when performing the LSSA fit, we allowed the free offset to be different in each run
\begin{equation}
A\sin(2 \pi f t) + B\cos(2 \pi f t) + \sum_i C_i\,\Pi_i(t) \, ,
\end{equation}
where $C_i$ is the free offset in the $i^\textrm{th}$ run and $\Pi_i(t)$ is a gate function equal to one in the $i^\textrm{th}$ run and zero elsewhere. This caused the short time--base analysis to lose sensitivity for periods longer than one run.
It should also be mentioned that, at the time of this analysis, the PSI data were still blinded, whereby an unknown, but constant, $d_\mathrm{n}$ was injected into them. It does not influence this analysis, as the free offsets are not considered further.

We split the $R$ time array into three sets: a control set of data without an applied electric field, and two sets sensitive to an oscillating EDM, namely with parallel and antiparallel applied electric and magnetic fields.
A coherent oscillating EDM signal would have an opposite phase in the latter two sets, and be absent in the control set. 
We did not perform a common fit. Instead, the two sensitive data sets were treated separately in the LSSA fits, and later combined to a limit. Otherwise, the LSSA treatment was the same as in the long time--base analysis. We picked a set of $156\,198$ trial frequencies, spaced apart at intervals determined by the spectral resolution (the inverse of 506 days = $\unit[23]{nHz}$), which here also defines the signal width. While we observed several peaks significant at the $3\sigma$ level (up to $6\sigma$), these did not have the correct characteristics in the three datasets to be consistent with the signal expected for axion DM.

%\begin{floatingfigure}[l]{0.5\textwidth}
%	\centering
%	\includegraphics[width=0.45\textwidth]{gfx/detection_psi_inset_gc.pdf}
%	\caption{ \footnotesize{Periodogram of the $R$ time array of the PSI experiment data, taken with the $\v{E}$ and $\v{B}$ fields parallel (black line). The mean of MC--generated periodograms, assuming no signal, is depicted in green. MC is used to calculate $1,2,…,5\,\sigma$ false--alarm thresholds, depicted in light orange. The periodogram of non-gradient-drift-corrected data is shown in pink.\hfill}}
%	\label{fig:PSI_detection}
%\end{floatingfigure}
%
%The periodogram of the $R$ time array taken with the parallel-field configuration is shown in Fig.\ref{fig:PSI_detection}. 
%The periodograms for the other two datasets (not shown) are very similar. 
%While we observe several peaks of significance of more than $3\sigma$, a non--statistical excess in a periodogram of $R$ may be caused not only by a coherent oscillating signal, but from a range of periodic effects too broad to fully account for.
%We defined strict requirements for an excess to be considered as one induced by axion DM as follows. 
%Firstly, the excess had to be observed in both sensitive datasets at the same frequency, but not in the control set. Secondly, the signals must %be in antiphase in the parallel and anti-parallel datasets. Lastly, we require high coherence (a narrow peak) equal to the spectral resolution of the dataset. None of the significant excesses passed our discovery criteria.

We delivered a limit on the oscillation amplitude similarly to the long time--base analysis, with the exception that we required the product of the two sensitive sets' $\mathrm{CL}_s$ statistics to be 0.05. 
The limit is shown as the blue curve in Fig.\,\ref{fig:ecm_limits}. 
Following Eq.\,(\ref{eq:nEDM_axion}), we can interpret the limit on the oscillating neutron EDM as limits on the axion--gluon coupling in Eq.\,(\ref{Axion_couplings}). 
We present these limits in Fig.\,\ref{fig:axion_limits_v2}, assuming that axions saturate the local cold DM energy density $\rho_{\rm DM}^{\rm local} \approx 0.4~\textrm{GeV/cm}^3$ \cite{Catena2010}.

%\FloatBarrier

\section{Axion-wind effect}
\begin{figure}
	\centering
	\includegraphics[width=0.45\textwidth]{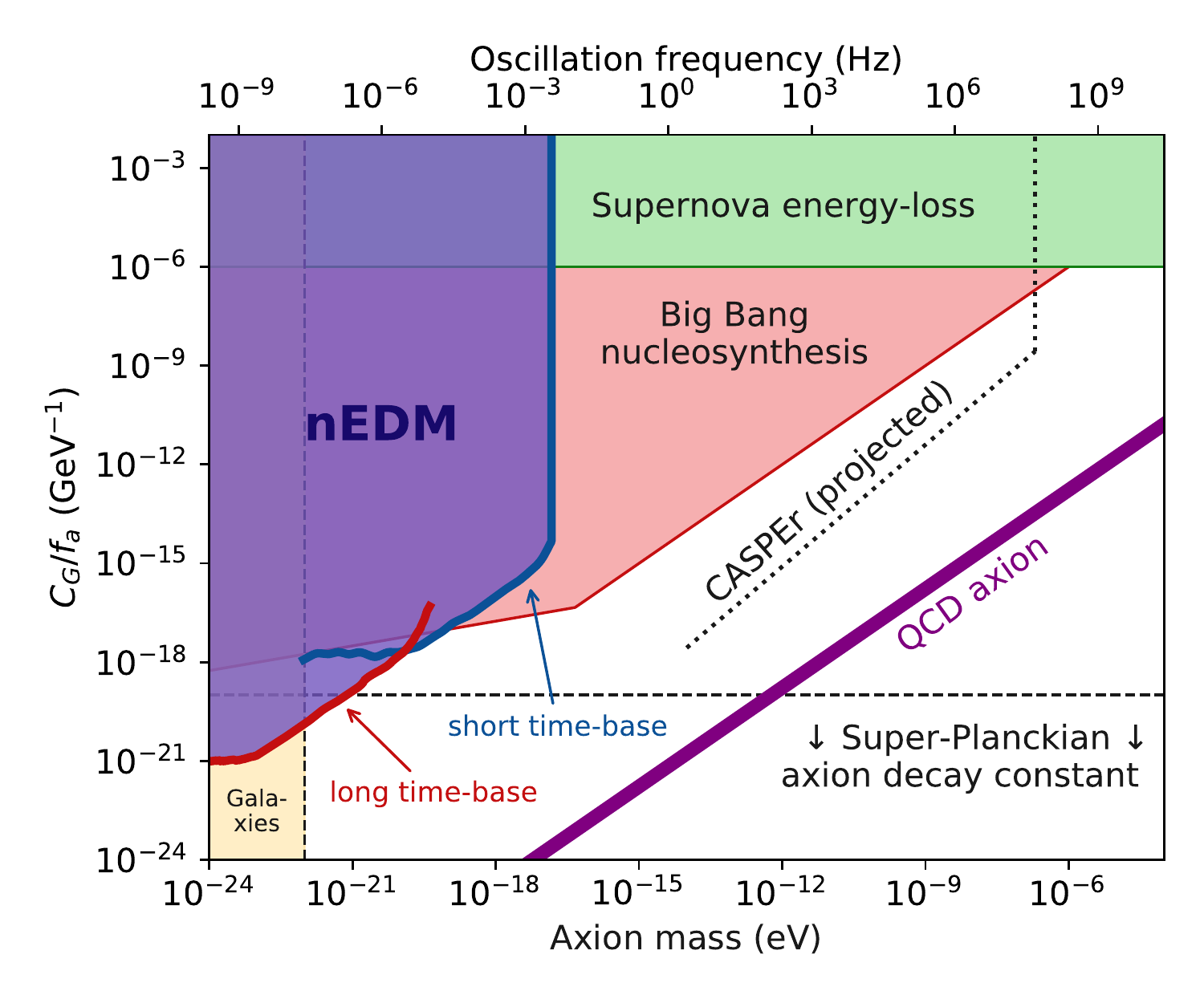}
	\qquad
	\includegraphics[width=0.45\textwidth]{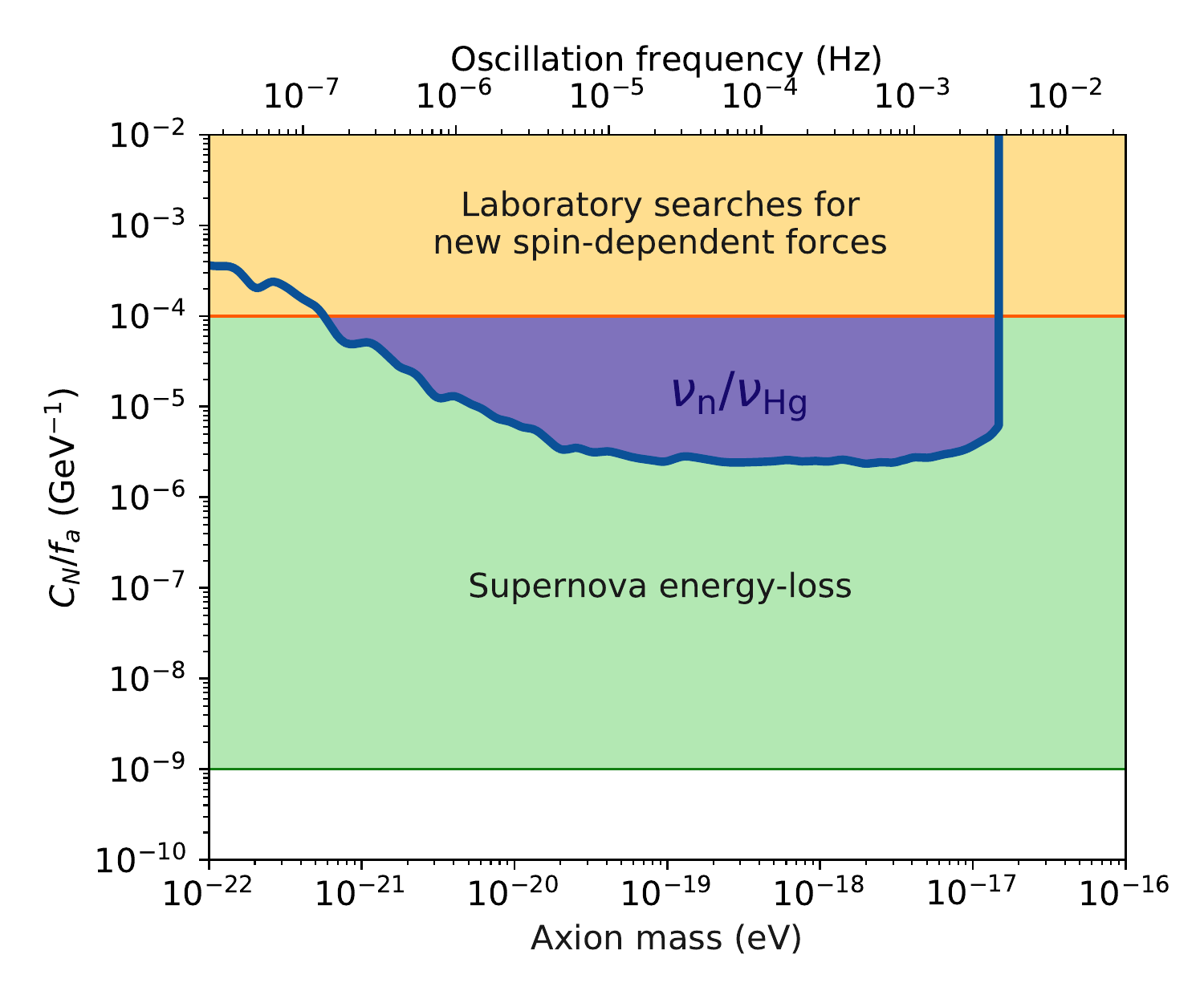}
	\caption{\footnotesize{Limits on the interactions of an axion with the gluons (left) and nucleons (right), as defined in Eq.\,(\ref{Axion_couplings}), assuming that axions saturate the local cold DM content. The regions above the thick blue and red lines correspond to the regions of parameters excluded by the present work at the 95\% confidence level (C.L.). The colored regions represent constraints from Big Bang nucleosynthesis (red, 95\% C.L.) \protect\cite{Blum2014,StadnikThesis,Stadnik2015D}, supernova energy-loss bounds (green, order of magnitude) \protect\cite{Graham2013,Raffelt1990Review,Raffelt2008LNP}, consistency with observations of galaxies (orange) \protect\cite{Marsh2015Review,Marsh2015B,Schive2015,Marsh2017}, and laboratory searches for new spin-dependent forces (yellow, 95\% C.L.) \protect\cite{Romalis2009_NF}. We also show the projected reach of the proposed CASPEr experiment (dotted black line)~\protect\cite{CASPEr2014}, and the parameter space for the canonical QCD axion (purple band). From \protect\cite{Abel2017}.}}
	\label{fig:axion_limits_v2}
\end{figure}

We also perform a search for the axion-wind effect, Eq.\,(\ref{potential_axion-wind}), by partitioning the entire PSI dataset into two sets with opposite magnetic-field orientations (irrespective of the electric field) and then analyzing the ratio $R = \nu_\mathrm{n} / \nu_\textrm{Hg}$ similarly to our oscillating EDM analysis above. The axion-wind signal would have an opposite phase in the two subsets. 
We find two overlapping $3\sigma$ excesses in the two subsets (at $\unit[3.42969]{\upmu Hz}$ and $\unit[3.32568]{mHz}$), neither of which have a phase relation consistent with an axion-wind signal. 
Following Eq.\,(\ref{potential_axion-wind}), we derive limits on the axion-nucleon coupling in Eq.\,(\ref{Axion_couplings}). 
We present these limits in Fig.\,\ref{fig:axion_limits_v2}, assuming that axions saturate the local cold DM energy density. 
Our peak sensitivity is 
$f_a/C_N \approx \unit[4 \times 10^{5}]{GeV}$ for $\unit[10^{-19}]{eV} \lesssim m_a \lesssim \unit[10^{-17}]{eV}$. 

\section{Conclusions}
In summary, we have performed a search for a time-oscillating neutron EDM in order to probe the interaction of axion-like dark matter with gluons. 
We have also performed a search for an axion-wind spin-precession effect in order to probe the interaction of axion-like dark matter with nucleons.
So far, no significant oscillations have been detected, allowing us to place limits on the strengths of such interactions. 
Our limits improve upon existing astrophysical limits on the axion-gluon coupling by up to 3 orders of magnitude and also improve upon existing laboratory limits on the axion-nucleon coupling by up to a factor of 40.
Furthermore, we constrain a region of axion masses that is complementary to proposed ``on-resonance'' experiments in ferroelectrics ~\cite{CASPEr2014}. Future EDM measurements will allow us to probe even feebler oscillations and for longer periods of oscillation that correspond to smaller axion masses.

\FloatBarrier
\pagebreak

\section*{References}
{%\footnotesize
	%\bibliography{bib}

\begin{thebibliography}{10}
	
	\bibitem{Abel2017}
	C.~Abel, {\it et~al.\/}, {\it Phys. Rev. X\/} {\bf 7}, 041034 (2017).
	
	\bibitem{Planck2015}
	P.~A.~R. Ade, {\it et~al.\/}, {\it Astron. Astrophys\/} {\bf 594}, A13 (2016).
	
	\bibitem{Axion-Review2015}
	P.~W. Graham, I.~G. Irastorza, S.~K. Lamoreaux, A.~Lindner, K.~A. van Bibber,
	{\it Annual Review of Nuclear and Particle Science\/} {\bf 65}, 485 (2015).
	
	\bibitem{Graham2011}
	P.~W. Graham, S.~Rajendran, {\it Phys. Rev. D\/} {\bf 84}, 055013 (2011).
	
	\bibitem{Stadnik2014A}
	Y.~V. Stadnik, V.~V. Flambaum, {\it Phys. Rev. D\/} {\bf 89}, 043522 (2014).
	
	\bibitem{Roberts2014A}
	B.~M. Roberts, {\it et~al.\/}, {\it Phys. Rev. Lett.\/} {\bf 113}, 081601
	(2014).
	
	\bibitem{Roberts2014B}
	B.~M. Roberts, {\it et~al.\/}, {\it Phys. Rev. D\/} {\bf 90}, 096005 (2014).
	
	\bibitem{Graham2013}
	P.~W. Graham, S.~Rajendran, {\it Phys. Rev. D\/} {\bf 88}, 035023 (2013).
	
	\bibitem{Witten1979}
	R.~Crewther, P.~D. Vecchia, G.~Veneziano, E.~Witten, {\it Physics Letters B\/}
	{\bf 88}, 123  (1979).
	
	\bibitem{Pospelov1999}
	M.~Pospelov, A.~Ritz, {\it Phys. Rev. Lett.\/} {\bf 83}, 2526 (1999).
	
	\bibitem{Catena2010}
	R.~Catena, P.~Ullio, {\it Journal of Cosmology and Astroparticle Physics\/}
	{\bf 2010}, 004 (2010).
	
	\bibitem{Altarev2009}
	I.~Altarev, {\it et~al.\/}, {\it Phys. Rev. Lett.\/} {\bf 103}, 081602 (2009).
	
	\bibitem{Altarev2010}
	I.~Altarev, {\it et~al.\/}, {\it EPL (Europhysics Letters)\/} {\bf 92}, 51001
	(2010).
	
	\bibitem{Afach2015_NF}
	S.~Afach, {\it et~al.\/}, {\it Physics Letters B\/} {\bf 745}, 58  (2015).
	
	\bibitem{Baker2014}
	C.~Baker, {\it et~al.\/}, {\it Nucl. Instr. Meth. Phys. Res. A\/} {\bf 736},
	184  (2014).
	
	\bibitem{Baker2011}
	C.~A. Baker, {\it et~al.\/}, {\it Physics Procedia\/} {\bf 17}, 159  (2011).
	
	\bibitem{Baker2006}
	C.~A. Baker, {\it et~al.\/}, {\it Phys. Rev. Lett.\/} {\bf 97}, 131801 (2006).
	
	\bibitem{Pendlebury2015}
	J.~M. Pendlebury, {\it et~al.\/}, {\it Phys. Rev. D\/} {\bf 92}, 092003 (2015).
	
	\bibitem{Ramsey1950}
	N.~F. Ramsey, {\it Phys. Rev.\/} {\bf 78}, 695 (1950).
	
	\bibitem{Steyerl1986}
	A.~Steyerl, {\it et~al.\/}, {\it Physics Letters A\/} {\bf 116}, 347  (1986).
	
	\bibitem{Scargle1982}
	J.~D. Scargle, {\it The Astrophysical Journal\/} {\bf 263}, 835 (1982).
	
	\bibitem{Cumming2004}
	A.~Cumming, {\it Monthly Notices of the Royal Astronomical Society\/} {\bf
		354}, 1165 (2004).
	
	\bibitem{Algeri2016}
	S.~Algeri, D.~van Dyk, J.~Conrad, B.~Anderson, {\it Journal of
		Instrumentation\/} {\bf 11}, P12010 (2016).
	
	\bibitem{PDG2016}
	C.~Patrignani, {\it et~al.\/}, {\it Chin. Phys.\/} {\bf C40}, 100001 (2016).
	
	\bibitem{Anghel2009}
	A.~Anghel, {\it et~al.\/}, {\it Nucl. Instr. Meth. Phys. Res. A\/} {\bf 611},
	272  (2009).
	
	\bibitem{Afach2015USSA}
	S.~Afach, {\it et~al.\/}, {\it European Physical Journal A\/} {\bf 51}, 1
	(2015).
	
	\bibitem{Ban2016NANOSC}
	G.~Ban, {\it et~al.\/}, {\it The European Physical Journal A\/} {\bf 52}, 326
	(2016).
	
	\bibitem{Knowles2009}
	P.~Knowles, {\it et~al.\/}, {\it Nucl. Instr. Meth. Phys. Res. A\/} {\bf 611},
	306  (2009). Particle Physics with Slow Neutrons.
	
	\bibitem{Afach2014magmoment}
	S.~Afach, {\it et~al.\/}, {\it Physics Letters B\/} {\bf 739}, 128  (2014).
	
	\bibitem{Afach2015}
	S.~Afach, {\it et~al.\/}, {\it Phys. Rev. Lett.\/} {\bf 115}, 162502 (2015).
	
	\bibitem{CASPEr2014}
	D.~Budker, P.~W. Graham, M.~Ledbetter, S.~Rajendran, A.~O. Sushkov, {\it Phys.
		Rev. X\/} {\bf 4}, 021030 (2014).
	
	\bibitem{Blum2014}
	K.~Blum, R.~T. D'Agnolo, M.~Lisanti, B.~R. Safdi, {\it Physics Letters B\/}
	{\bf 737}, 30  (2014).
	
	\bibitem{StadnikThesis}
	Y.~V. Stadnik. \emph{Manifestations of Dark Matter and Variations of the
		Fundamental Constants of Nature in Atoms and Astrophysical Phenomena},
	(Springer, Cham, 2017).
	
	\bibitem{Stadnik2015D}
	Y.~V. Stadnik, V.~V. Flambaum, {\it Phys. Rev. Lett.\/} {\bf 115}, 201301
	(2015).
	
	\bibitem{Raffelt1990Review}
	G.~G. Raffelt, {\it Physics Reports\/} {\bf 198}, 1  (1990).
	
	\bibitem{Raffelt2008LNP}
	G.~G. Raffelt, {\it Astrophysical Axion Bounds\/} (Springer Berlin Heidelberg,
	Berlin, Heidelberg, 2008), pp. 51--71.
	
	\bibitem{Marsh2015Review}
	D.~J. Marsh, {\it Physics Reports\/} {\bf 643}, 1  (2016).
	
	\bibitem{Marsh2015B}
	B.~Bozek, D.~J.~E. Marsh, J.~Silk, R.~F.~G. Wyse, {\it Monthly Notices of the
		Royal Astronomical Society\/} {\bf 450}, 209 (2015).
	
	\bibitem{Schive2015}
	H.-Y. Schive, T.~Chiueh, T.~Broadhurst, K.-W. Huang, {\it The Astrophysical
		Journal\/} {\bf 818}, 89 (2016).
	
	\bibitem{Marsh2017}
	P.~S. Corasaniti, S.~Agarwal, D.~J.~E. Marsh, S.~Das, {\it Phys. Rev. D\/} {\bf
		95}, 083512 (2017).
	
	\bibitem{Romalis2009_NF}
	G.~Vasilakis, J.~M. Brown, T.~W. Kornack, M.~V. Romalis, {\it Phys. Rev.
		Lett.\/} {\bf 103}, 261801 (2009).
	
\end{thebibliography}

}

\end{document}